\documentclass[ usenatbib]{mn2e}
\usepackage{epsfig}
\usepackage{amsmath} 

\newcommand{\cm}{{\rm cm}$^{-2}$}
\newcommand{\cmt}{{\rm cm}$^{-3}$}
\newcommand{\tb}{{\rm K \rm km \rm s}$^{-1}$}
\newcommand{\msun}{{\rm M}$_{\odot}$}

\newcommand{\myr}{\rm Myr}

\title{Using CO line ratios to trace the physical properties of molecular clouds }
\author[C.H.Pe\~naloza et al.]{Camilo H. Pe\~naloza$^{1}$, Paul C. Clark$^{1}$, Simon C.O. Glover$^{2}$,
\and Rahul Shetty$^{2}$, \& Ralf S. Klessen$^{3}$ \\
$^{1}$School of Physics and Astronomy, The Parade, Cardiff University, Cardiff CF24 3AA, UK\\	
$^{2}$Universit\"at Heidelberg, Zentrum f\"ur Astronomie, Institut f\"ur Theoretische Astrophysik, Albert-Ueberle-Str. 2, 69120
Heidelberg, Germany\\
$^3$Universit\"{a}t Heidelberg, Interdiszipli\"{a}res Zentrum f\"{u}r Wissenschaftliches Rechnen, Heidelberg, Germany 
}

\begin{document}

\maketitle

\begin{abstract}

The carbon monoxide (CO) rotational transition lines are the most common tracers of molecular gas within giant molecular clouds (MCs). We study the ratio ($R_{2-1/1-0}$) between CO's first two emission lines and examine what information it provides about the physical properties of the cloud. To study $R_{2-1/1-0}$ we perform smooth particle hydrodynamic simulations with time dependent chemistry (using GADGET-2), along with post-process radiative transfer calculations on an adaptive grid (using RADMC-3D) to create synthetic emission maps of a MC. $R_{2-1/1-0}$ has a bimodal distribution that is a consequence of the excitation properties of each line, given that $J=1$ reaches local thermal equilibrium (LTE) while $J=2$ is still sub-thermally excited in the considered clouds. The bimodality of $R_{2-1/1-0}$ serves as a tracer of the physical properties of different regions of the cloud and it helps constrain local temperatures, densities and opacities. Additionally this bimodal structure shows an important portion of the CO emission comes from diffuse regions of the cloud, suggesting that the commonly used conversion factor of $R_{2-1/1-0}\sim 0.7$ between both lines may need to be studied further. 

\end{abstract}


\section{Introduction} \label{introsec}

The characterisation of giant molecular clouds (GMCs) -- the sites of nearly all star formation activity in the local Universe -- is an important step towards understanding how stars are born.  Molecular hydrogen (H$_2$) is the most abundant molecule in the interstellar medium (ISM), but its rotational emission lines are not excitable at the temperatures found in most GMCs. However, the second most abundant molecule is carbon monoxide (CO), which has rotational transitions that are easily excitable within typical GMCs, making CO a good tracer of molecular gas. Additionally the lower levels of CO emit at a frequency that can be observed from the ground. Therefore, CO has become the favoured tracer for studying molecular gas in GMCs \citep{Liszt:1998tx, Dame:2001bg,McKee:2007bd,2008ApJ...680..428G,Bolatto:2013hl,2016SAAS...43...85K}.  

CO is not without problems as a tracer of molecular gas. Its emission is highly sensitive to environmental conditions \citep{2012A&A...541A..58L}, and traces only a limited range of column densities. At low column densities, CO is rapidly photodissociated by the interstellar radiation field (ISRF) \citep{1988ApJ...334..771V}, while at high column densities, its emission lines become optically thick. For the $J = 1-0$ line of $^{12}$CO, this occurs at a CO column density of $N_{\rm CO} \approx 10^{16}$\cm, corresponding to a visual extinction of only a few \citep{Liszt:1998tx}. Observations of this line therefore do not directly probe the highest density regions of the cloud.

Despite this, CO emission still contains plenty of information about the cloud's conditions and structure. A study by \citet{1990A&A...234..469C} illustrates this by investigating the emission ratio ($R_{2-1/1-0}$) for CO's lowest two rotation transitions, $J = 2-1$ and $J=1-0$. This ratio is conventionally defined as 
\begin{equation}
R_{2-1/1-0}=\frac{W_{2-1}}{W_{1-0}},  
\label{eq:ratio}
\end{equation}
where $W_{1-0}$ and $W_{2-1}$ are the velocity-integrated brightness temperatures of the $J=1-0$ and $J=2-1$ rotational transition lines of CO, expressed in units of ${\rm K \, km \, s^{-1}}$. These two transitions have energy separations $E_{10} / k_{\rm B} = 5.5$~K and $E_{21} / k_{\rm B} = 11.04$~K, respectively, where $k_{\rm B}$ is Boltzmann's constant. Their critical densities in fully molecular gas with a temperature $T = 10$~K are $n_{\rm crit, 1-0} \simeq 2000 \: {\rm cm^{-3}}$ and $n_{\rm crit, 2-1} \simeq 10000 \: {\rm cm^{-3}}$. As most of the gas in a molecular cloud has a density $n_{\rm H_{2}} < n_{\rm crit, 2-1}$, the value of $R_{2-1/1-0}$ is sensitive to both the density and the temperature structure of the gas, as well as the optical depth of the two lines. 

The behaviour of $R_{2-1/1-0}$ on small scales within molecular clouds has been examined by  \citet{1994ApJ...425..641S} and more recently by \citet{2015ApJS..216...18N}. They studied how the line ratio varies within the Orion GMC, finding that $R_{2-1/1-0} \sim 1$ towards the centre of the cloud, but that it declines towards the outskirts where $R_{2-1/1-0} \sim 0.5$. \citet{1994ApJ...425..641S} argue that the observed variations can be understood as a consequence of the density variations within the cloud. This is a reasonable assumption if the CO-emitting gas is isothermal, but we know from numerical simulations of molecular clouds that this is only approximately true and that temperature variations of a factor of two or more within CO-rich gas are not uncommon \citep[see e.g.][]{Glover:2010bu}. Further complicating matters is the fact that the variations in density and temperature are not independent: the density structure depends sensitively on the temperature of the gas, while the temperature depends both on the density, and also on other factors such as the local extinction, the metallicity of the gas and the strength of the ISRF.

In order to better understand what the CO line ratio can tell us about the physics of the cloud, we make use of numerical models which satisfactorily reproduce the irregular structure of the gas. This has become practical within the last few years with the advent of 3D dynamical models of GMCs that account for the chemical and thermal evolution of the gas, the non-isotropic nature of the attenuated radiation field, and the complex morphology of the cloud whilst still being computationally reasonable (see e.g.\ \citealt{Glover:2010bu}, \citeauthor{Clark_etal_2012}~2012a, \citealt{2012MNRAS.427.2100B}, \citealt{Offner:2013du}).

In this paper, we make use of these techniques to study the behaviour of $R_{2-1/1-0}$ within a turbulent molecular cloud. We perform a 3D hydrodynamical simulation of a representative cloud that self-consistently follow the thermal and chemical evolution of the gas. We then post-process the results of this simulation to generate synthetic $^{12}$CO 1-0 and 2-1 emission maps.\footnote{From this point on, when we refer to CO, we mean $^{12}$CO, unless otherwise noted.}
 The resulting maps allow us to study in detail the relationship between the line ratio and the physical conditions in the cloud.

The structure of the paper is as follows. In Section \ref{metsec}, we outline our method for modelling a molecular cloud and also describe how we post-process the simulations to generate synthetic emission maps.  Section \ref{ressec} presents the results from our simulations and the analysis of the emission lines and $R_{2-1/1-0}$. Section \ref{dissec} discusses possible explanations for the physical processes driving the behaviour $R_{2-1/1-0}$ and how this is consistent with our findings. Finally we summarise all of our findings in Section \ref{consec}.

\section{Method}\label{metsec}

\subsection{Numerical setup}

\subsubsection{Hydrodynamics and chemistry}

To model the gas in this study, we use a modified version of the publicly available smoothed particle hydrodynamics (SPH) code, GADGET-2 \citep{Springel:2005cz}. The changes include a time-dependent chemical network that follows the formation and destruction of H$_2$  \citep{Glover:2007gr,Glover:2007wq} and CO \citep{NelsonLanger1999}, more details of which can be found in \citet{Glover:2012et}, which also includes the photodissociation rates that we adopt in this study.  We adopt the same radiative heating and cooling rates, and cosmic ray heating rate as described in \citet{Glover:2007gr} and \citet{Glover:2012dd}. To treat the attenuation of the ISRF we use the {\sc TreeCol} algorithm developed by \citeauthor{Clark_etal_2012}~(2012a).
\subsubsection{Initial conditions}
\label{sec:ics}

Our initial setup uses a $10^4$ \msun\ uniform sphere  $(R\sim8.84\ {\rm pc})$, with an initial volume density of $n=100$ \cmt\ ($n$ is given for a mean molecular weight of $\mu = 1.4 $) and $2\times 10^6$ SPH particles. We impose a turbulent velocity field with a power spectrum of $P(k) \propto k^{-4}$, in which the energy is partitioned into a natural mixture of solenoidal and compressive modes. The energy in the turbulent velocity field is set such that $E_{\rm pot} / E_{ \rm kin}  = \epsilon = 2$  (i.e.\ the cloud is gravitationally bound). This kinetic energy is allowed to decay freely via shock dissipation.

We adopt solar metallicity (${\rm Z} = {\rm Z}_{\odot}$), and a standard dust-to-gas ratio of 0.01. For the ISRF, we use a spectral shape taken from \citet{Draine1978} at ultraviolet wavelengths and  \citet{Black1994} at longer wavelengths. The strength of the ISRF is scaled such that $G_{0} = 1.7$ in \citet{Habing1968} units, where $G_{0}$ is the energy density in the range 6--13.6~eV. At the beginning of the simulation, hydrogen is assumed to be fully molecular, i.e.\ $f({\rm H}_2)=1$, oxygen is in its atomic form, and carbon is assumed to be in the form of C$^+$. 

\citeauthor{Clark2012b}~(2012b) demonstrated that the H$_2$ fraction rises sharply to near unity in the compression events that form molecular clouds.  However, it has also been shown that the initial chemical state of the cloud has little effect on the global evolution \citep{Glover:2012et,Glover:2012dd,2015MNRAS.452.2057C}. In this study, we analysed our results for clouds which started both fully atomic and fully molecular, finding no significant difference. In the interest of clarity, we present only the results from the clouds with $f({\rm H}_2) = 1$ initially. 

As we are interested in the properties of the gas, and not the star formation that takes place inside the cloud,  we stop the simulation just as the collapse of the first pre-stellar core occurs. This takes place at about $1.91$ \myr\ for our simulated cloud. At this point, we produce a snapshot containing the positions, velocities, temperatures, dust densities and molecular number densities for each SPH particle. This snapshot contains the necessary data to perform radiative transfer simulations and produce synthetic emission maps. 

\subsection{Post-processing}

\subsubsection{Radiative transfer simulations}

To produce synthetic observations of the CO emission, we need to post-process our final simulation snapshot with a line radiative transfer code. In this study we use the publicly available radiative transfer code RADMC-3D \citep{2012ascl.soft02015D}. The high optical depth of CO means that the populations in the first and second energy levels are often close to those expected for molecules in local thermal equilibrium (LTE). However, this is not always the case, in particular in gas that has a low density or low optical depth. Therefore, we use the large velocity gradient (LVG) approximation \citep{1957SvA.....1..678S} to account for the non-LTE level populations in these regions. A detailed description of the implementation of the LVG algorithm in RADMC-3D can be found in \citet{Shetty:2011eh}. 

For the level population calculations, RADMC-3D requires the number density of CO, the number density of its dominant collision partner H$_{2}$,  the temperature and the velocity of the gas, all which come directly from our hydrodynamic simulation. Additionally the molecular properties for CO are taken from the Leiden Atomic and Molecular Database \citep{2005A&A...432..369S}. The collisional excitation rates that we adopt come originally from the study of \citet{2010ApJ...718.1062Y}. Finally, we include a microturbulence velocity dispersion of $v=0.2 \, {\rm km s}^{-1}$ to account for small-scale broadening of the spectral lines by unresolved velocity fluctuations. The magnitude of this microturbulent velocity is chosen to be consistent with the \citet{1981MNRAS.194..809L} size-linewidth relation.

\subsubsection{Grid interpolation} \label{gridsubsec}

To post-process the SPH data in RADMC-3D, one first needs to map the unstructured SPH particle distribution onto a Cartesian grid. Interpolation onto a uniform cartesian grid (see e.g.\ \citealt{Glover:2012jo} and \citealt{2014MNRAS.445.4055S}) is straightforward, but has the limitation that it is not well suited to account for the varying spatial resolution that exists in GADGET-2's particle distribution. In high density regions, the Lagrangian nature of SPH means that the particles are closely spaced, but this information can be lost if they are interpolated onto a grid with a cell size that is larger than the inter-particle spacing. The obvious solution to this problem is to require the cell size to be smaller than the smallest particle separation, but to achieve this with a uniform grid is computationally infeasible and would require a grid resolution of around $4096^{3}$ for the simulation we present here.

In our present study, we therefore make use of an alternative solution. RADMC-3D is capable of constructing and utilising oct-tree grids (similar to those used in some adaptive mesh refinement codes, such as FLASH; see e.g.\ \citealt{Fryxell00}), and this structure is a much more natural fit to the disordered SPH particle distribution. We therefore interpolate the data from the SPH particles onto a suitably-constructed oct-tree grid, ensuring that no data is lost during the interpolation process. Full details of our methodology can be found in Appendix~\ref{appsec}.

\begin{figure*}
\includegraphics[width=\textwidth]{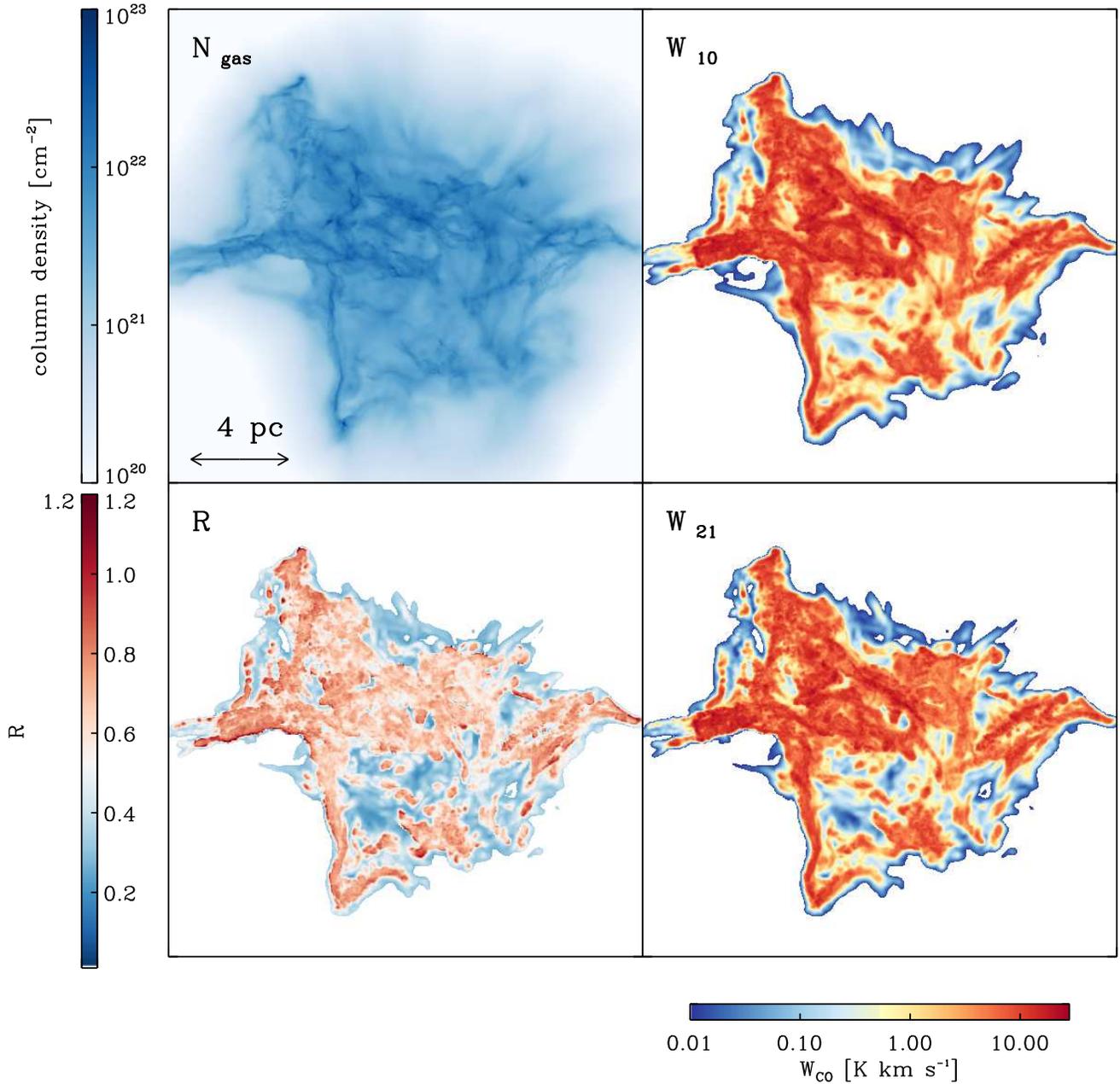}
\caption{Top left: Column density of hydrogen nuclei, $N_{\rm gas}$, at the end of the simulation. Since the gas in the cloud is primarily molecular, the H$_{2}$ column density is given approximately by $N_{\rm H_{2}} \simeq N_{\rm gas} / 2$.
Top right and bottom right: the velocity-integrated intensity of the cloud at the same time, for the J$=1-0$ and J$=2-1$ emission lines, respectively. Bottom left: the emission line ratio $R_{2-1/1-0}$, as defined in  Equation~\ref{eq:ratio}.}
\label{fig:MAPS}
\end{figure*}

\section{The CO 2-1 / CO 1-0 line ratio}\label{ressec}

\subsection{CO emission maps}
 
The column density map in the upper-left panel of Fig.~\ref{fig:MAPS} gives an overview of the gas distribution and density of the cloud at the end of the simulation. The column density shown is the total column density of H nuclei, $N_{\rm gas} = N_{\rm H} + N_{\rm H^{+}} + 2 N_{\rm H_{2}}$. However, as the gas is predominantly molecular, we find that in practice, $N_{\rm gas} \simeq 2 N_{\rm H_{2}}$.The column density map clearly shows the filamentary structure produced in the cloud by turbulence and self-gravity.

Maps of the velocity-integrated intensity of the CO 1-0 and 2-1 emission lines, $W_{10}$ and $W_{21}$, are also shown in Fig.~\ref{fig:MAPS}, in the upper-right and lower-right panels, respectively. Comparing the column density map and the integrated intensities for both lines, we see that the bulk of the gas in the cloud is well-traced by the emission. Many of the filamentary structures visible in the column density map are flattened out due to line saturation. However, their locations are still visible in the maps of $W_{10}$ and $W_{21}$. Towards the denser regions of the filaments, the emission is much brighter and more structure can be observed. Note, however, that the change in the column density as we move from one of the filaments to the surrounding gas is much smaller than the corresponding change in the CO integrated intensity; much of the CO in the lower density gas is photodissociated by the ISRF.

Comparing $W_{10}$ and $W_{21}$, we see that both maps show similar structure, with the most obvious difference being that in general$W_{10}$ is slightly brighter than $W_{21}$. This is particularly apparent towards the centre of the filaments, or at the outskirts of the cloud, where the gas is most diffuse. Nevertheless, the overall integrated intensity of both lines is very similar.

\subsection{The value of $R_{2-1/1-0}$}

\begin{figure}
\centering
\includegraphics[width=1.\linewidth]{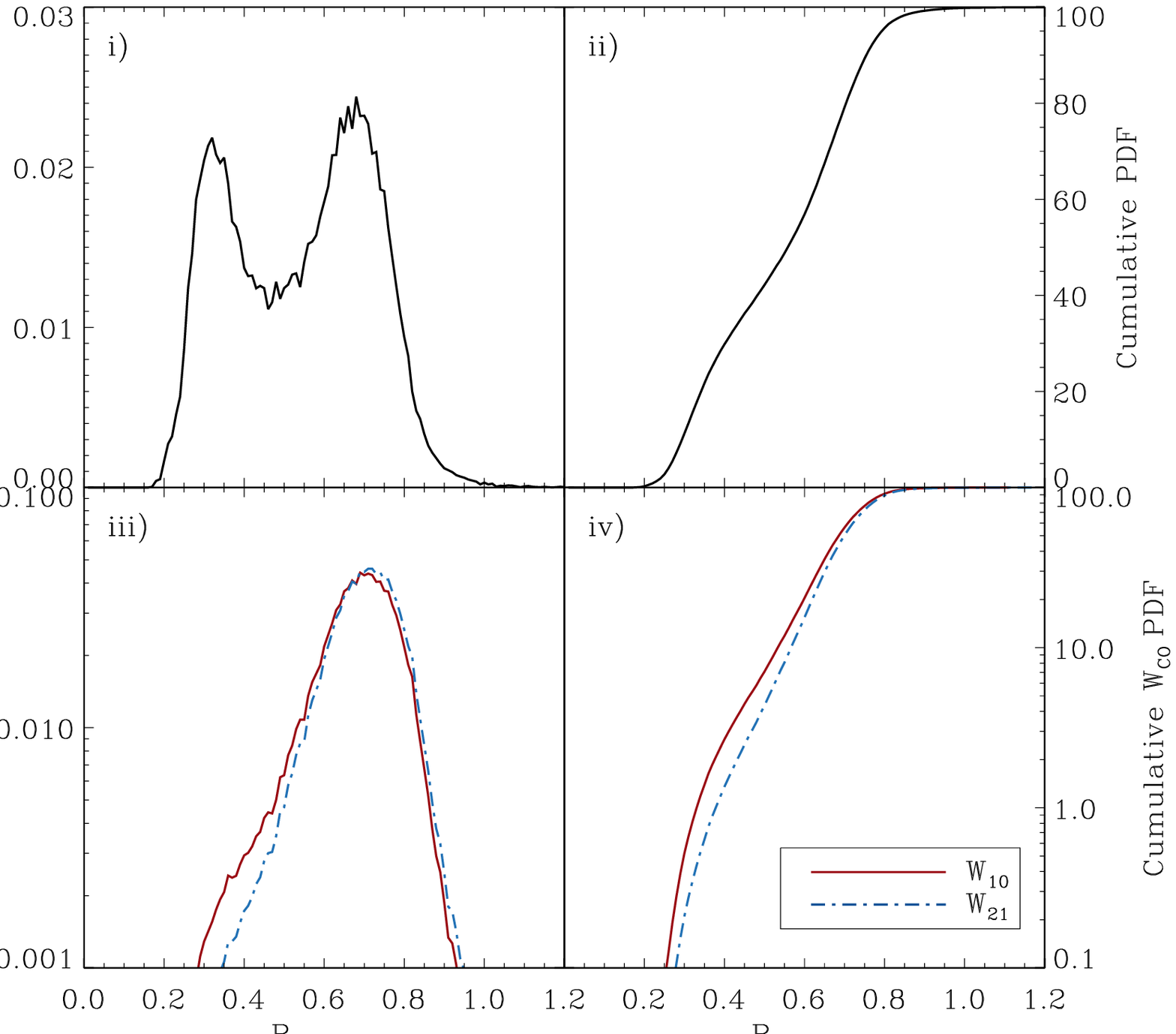}
\caption{(i) Probability density function (PDF) of $R_{2-1/1-0}$, illustrating its bimodal behaviour. (ii) Cumulative PDF of $R_{2-1/1-0}$. (iii) PDF of $R_{2-1/1-0}$, weighted by the integrated brightness temperature $W_{\rm CO}$ of the 1-0 line (red solid) or the 2-1 line (blue dashed-dotted). (iv) Cumulative version of (iii).} 
\label{fig:Histograms}
\end{figure}

The images in Fig. \ref{fig:MAPS} discussed above show that the emission from both the CO 1-0 and CO 2-1 lines is very similar. In order to highlight the differences that do exist, we compute the ratio $R_{2-1/1-0}$, as defined in Equation \ref{eq:ratio}, for each pixel in the synthetic image. The resulting distribution of intensity ratios is shown in the bottom left panel of Fig. \ref{fig:MAPS}. This figure suggests that the distribution of $R_{2-1/1-0}$ could be bimodal.

To better quantify the variations in the line ratio, we construct a probability density function (PDF) for $R_{2-1/1-0}$, which is shown in panel (i) of Fig.~\ref{fig:Histograms}. This PDF is constructed using area weighting, meaning that each pixel in the synthetic images is weighted equally. The figure confirms the 
bimodal behaviour of $R_{2-1/1-0}$: there are two distinct peaks, one centred at $R_{2-1/1-0}\sim 0.7$ and the other at $R_{2-1/1-0}$ $\sim 0.3$. If we consider the cumulative PDF, as shown in Panel (ii) of Fig. \ref{fig:Histograms}, then we see that the high ratio peak represents around 60\% of the total cloud area, and the low ratio peak represents the remaining 40\%. 

Observationally, measurements of $R_{2-1/1-0}$ for real molecular clouds or collections of molecular clouds typically recover values similar to those that we find for the high ratio peak. For example, \citet{1994ApJ...425..641S} report mean values of $R_{2-1/1-0}$ of 0.77 for the Orion A molecular cloud and 0.66 for the Orion B cloud. On larger scales, \citet{Koda:2012ht} report values of around 0.7--0.9 for clouds in the spiral arms of M51, although for the inter-arm clouds they find somewhat lower values of 0.4--0.6. It is also worth noting that a common value adopted in the literature for converting from $W_{21}$ to $W_{10}$ is $R_{2-1/1-0}\sim0.7$ \citep{1990ApJ...348..434E,1991A&A...251....1C,1995A&A...303..851B,1997ApJ...486..276S,1997IAUS..170...39H,2001ApJS..136..189S,2008AJ....136.2846B,2009AJ....137.4670L,2011MNRAS.416.1250B}. 

The reason why these previous studies have not detected or discussed the lower ratio peak becomes clear when we examine the emission-weighted PDF of $R_{2-1/1-0}$, shown in panel (iii) of Fig.~\ref{fig:Histograms}. It is evident from this plot that although both peaks in the area-weighted PDF correspond to similar areas, they correspond to very different total intensities. The high ratio peak corresponds to regions of the cloud that are bright in CO, and hence shows up clearly in the emission-weighted PDF. The low ratio peak, however, is produced by emission from regions with very low CO brightness, and hence essentially disappears in the emission-weighted PDF, remaining visible only as a small wing on the left-hand side of the main peak. We see also that we recover the same behaviour regardless of whether we weight the PDF using the integrated intensity of the (1-0) transition (the red curve in panel (iii) of Fig.~\ref{fig:Histograms}) or the (2-1) transition (the blue curve in the same Figure).

\begin{figure}
\centering
\includegraphics[width=\linewidth]{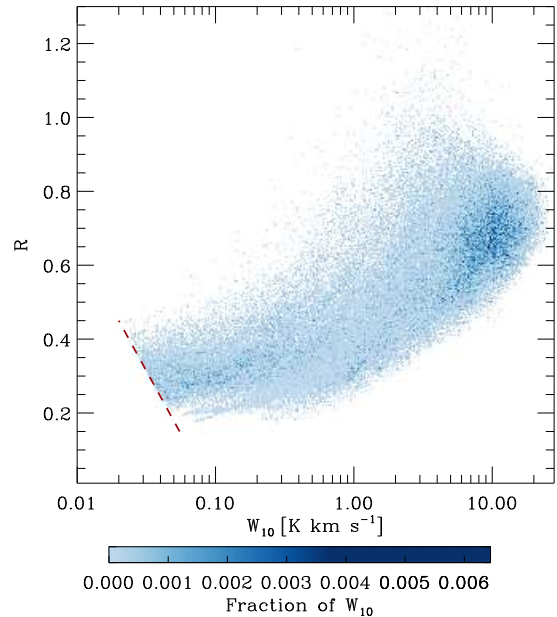}
\caption{
$R_{2-1/1-0}$, plotted as a function of the integrated intensity of the CO 1-0 line, $W_{10}$. Values are plotted for all pixels in the synthetic emission
maps that have $W_{10} > 0.01 \: {\rm K \, km s^{-1}}$ and $W_{21} > 0.01 \: {\rm K \, km s^{-1}}$. The diagonal red dashed line indicate this selection criterion. 
Note that in practice, $R < 0.4$ when $W_{10} < 0.1 \: {\rm K \, km s^{-1}}$, so a number of points are removed that have $W_{21} < 0.01 \: {\rm K \, km s^{-1}}$
but $W_{10} > 0.01 \: {\rm K \, km s^{-1}}$.}
\label{fig:TB2peak}
\end{figure}

To get a feel for the sensitivity that would be required to observe $R_{2-1/1-0}$, in Fig.~\ref{fig:TB2peak} we show how $R_{2-1/1-0}$ varies as a function of the integrated intensity of the 1-0 line. This plot again demonstrates that lines-of-sight with low values of $R_{2-1/1-0}$ have low integrated intensities. For example, essentially all of the lines-of-sight that have $R_{2-1/1-0} \sim 0.3$ have CO 1-0 integrated intensities that are less than 1 K km$\,{\rm s}^{-1}$. Observations with sensitivities of $\ga 1$~K km$\,{\rm s}^{-1}$ will therefore simply not detect the emission from these regions.

To put these values into context, note that the CO (1-0) map used in the \citet{1994ApJ...425..641S} study of Orion A and B (taken originally from \citealt{Maddalena:1986}) has a brightness temperature sensitivity of around 0.8~K, while the CO (2-1) map made by \citet{1994ApJ...425..641S} themselves has roughly a factor of two better sensitivity. If we assume a cloud velocity width of a few km~s$^{-1}$, this corresponds to a minimum integrated intensity of a few K~km~s$^{-1}$ for both lines. It is therefore unsurprising that they recover only high values for $R_{2-1/1-0}$.  Interestingly, the more recent study of Orion A and B by \citet{2015ApJS..216...18N}, which had a $3\sigma$ sensitivity of $0.24$ \tb, recovers values of $R_{2-1/1-0} \sim 0.4$ or lower in some regions of the cloud (particularly towards the left side of the ridge, away from the OB association),  consistent with our argument above that high sensitivity is required in order to observe regions with low $R_{2-1/1-0}$.

\subsection{The dependence of $R_{2-1/1-0}$ on density and temperature}

\begin{figure}
\centering
\includegraphics[width=1\linewidth]{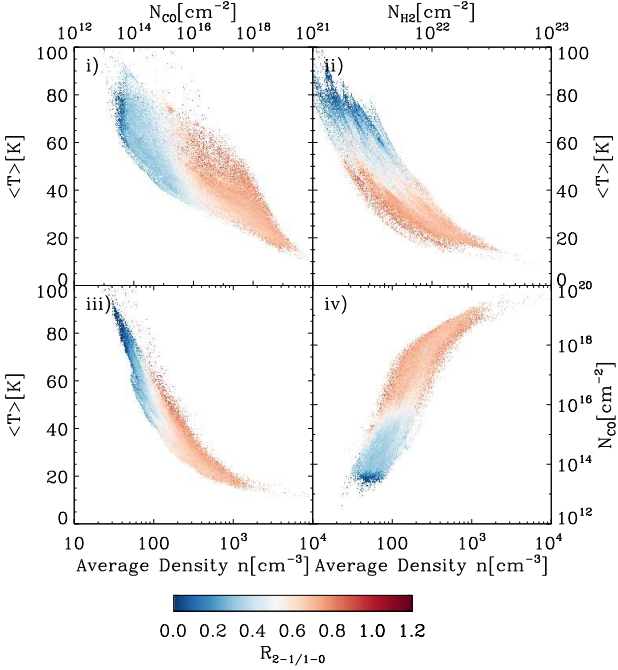}
\caption {Different physical quantities are plotted and colour coded with $R_{2-1/1-0}$: (i) N$_{\rm CO}$ vs $\langle T \rangle$; (ii) N$_{\rm H_2}$ vs $\langle T \rangle$; (iii) $\langle n \rangle$ vs $\langle T \rangle$; (iv) $\langle n \rangle$ vs N$_{\rm CO}$.}
\label{fig:quantities}
\end{figure}

Considering the distribution of $R_{2-1/1-0}$ in out maps in Fig. \ref{fig:MAPS}, it is worth exploring how it relates to physical quantities within the cloud such as temperature or density. The bimodal behaviour we see in the area-weighted PDF suggests that the local conditions of the cloud are changing such that emission from one or both lines is affected, creating two peaks in $R_{2-1/1-0}$. We investigate how $R_{2-1/1-0}$ varies as a function of a number of different quantities computed for each line of sight: the mean temperature $\langle T \rangle = \sum_{i} T_i m_i$, the mean number density  $\langle n \rangle$, or the H$_{2}$ and CO column densities (N$_{\rm H_2}$ and N$_{\rm CO}$). We also examine how these quantities correlate with each other. The results are shown in~Fig. \ref{fig:quantities}. Note that the colour map used in these plots to indicate $R_{2-1/1-0}$ is the same as that of Fig.~\ref{fig:MAPS}. 

Panel (i) in Fig.~\ref{fig:quantities} shows N$_{\rm CO}$ plotted against $\langle T \rangle$. Although there is a clear inverse correlation between these two quantities, there is also significant scatter in the mean temperature associated with any given CO column density. This is a consequence of the fact that other than at the very highest CO column densities, warm, CO-poor gas makes a significant contribution to $\langle T \rangle$ but has little influence on N$_{\rm CO}$. This is not surprising since this follows from the heating and cooling processes of the gas \citep{1965ApJ...142..531F,1969ApJ...155L.149F, Glover:2012et}. Therefore two sight-lines that probe similar amounts of CO but differing amounts of warm gas, can have quite different mean temperatures associated with the same CO column density. Consequently, $\langle T \rangle$ is only a good measure of the temperature of the CO-emitting gas when the CO column density is large. We also see that $R_{2-1/1-0}$ has a strong dependence on N$_{\rm CO}$: there is a clear rapid shift at N$_{\rm CO}\sim 10^{15}$ \cm\ separating gas with low $R_{2-1/1-0}$ from gas with high $R_{2-1/1-0}$. On the other hand, $R_{2-1/1-0}$ depends only weakly on $\langle T \rangle$, largely because $\langle T \rangle$  in general is not a good measure of the temperature of the CO-emitting gas.

Panel (ii) in Fig.~\ref{fig:quantities} depicts N$_{\rm H_2}$ plotted against $\langle T \rangle$. Again, there is a clear inverse correlation, reflecting the fact that lines-of-sight with high H$_{2}$ column densities preferentially sample dense gas that is well-shielded from the ISRF and that hence is cold. Looking at the behaviour of $R_{2-1/1-0}$ in this plot, we see that although it is low when N$_{\rm H_{2}}$ is small (N$_{\rm H_{2}} \sim 10^{21} \: {\rm cm^{-2}}$)  and high when N$_{\rm H_{2}}$ is large (N$_{\rm H_{2}} > 10^{22} \: {\rm cm^{-2}}$), at intermediate column densities there is no clear correlation between $R_{2-1/1-0}$  and N$_{\rm H_{2}}$. 

In panel (iii) of Fig.~\ref{fig:quantities}, we illustrate the relationship between $\langle n \rangle$ and $\langle T \rangle$. In this case, there is a relatively tight inverse correlation, showing that lines-of-sight with low mean density probe primarily warm gas, while lines-of-sight with high mean density probe cold gas. Once again, there is a clear bimodality in the behaviour of $R_{2-1/1-0}$: low values correlate well with low mean densities and high mean temperatures, while high values correlate with high mean densities and low mean temperatures. 

Finally, in panel (iv) of Fig.~\ref{fig:quantities}, we present $\langle n \rangle$ against N$_{\rm CO}$.  We see from this plot that although there is a clear correlation between the mean density along a sight-line and the CO column density of that sight-line, there is a substantial scatter in this relationship for values of $\langle n \rangle$ around $\langle n \rangle \sim 100 \: {\rm cm^{-3}}$. As  $R_{2-1/1-0}$ correlates more strongly with N$_{\rm CO}$ than with $\langle n \rangle$, the result is that there is only a weak relationship between the mean density and the value of $R_{2-1/1-0}$ for mean densities close to $100 \: {\rm cm^{-3}}$. However, it is also clear that $R_{2-1/1-0}$ is always large when $n \gg 100 \: {\rm cm^{-3}}$, and always small if $n \ll 100 \: {\rm cm^{-3}}$.  

Putting this all together, we see that there are clear links between the bimodal structure visible in the distribution of $R_{2-1/1-0}$ and the mean values of the physical conditions (density, temperature, etc.) within the cloud. Lines-of-sight with high $R_{2-1/1-0}$ preferentially probe regions with high CO column densities, high mean densities and low temperatures. Conversely, lines-of-sight with low $R_{2-1/1-0}$ probe regions with low CO column densities, low mean densities and high mean temperatures. However, these relationships do not explain why the transition from low $R_{2-1/1-0}$ to high $R_{2-1/1-0}$ occurs so suddenly, or why it is the CO column density in particular that best predicts when this transition will occur. To understand why this happens, we need to look at how the CO line opacities vary within the cloud.

\subsection{Optical depth effects}

\begin{figure}
\includegraphics[width=1\linewidth]{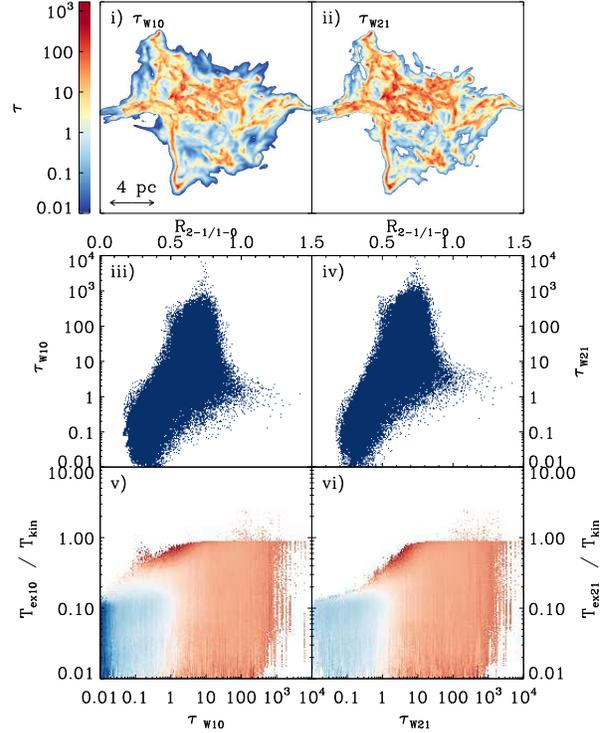}
\caption{Panels i) and ii) show Integrated opacity image weighted by integrated intensity for both lines ($\tau_{W10}$ and $\tau_{W21}$ respectively). Panels iii) and iv) show $\tau_{W}$, plotted as a function of $R_{2-1/1-0}$ for both lines illustrating how the ratio and opacity are correlated. Panels v) and vi) show T$_{\rm ex}$/T$_{\rm kin}$ plotted as a function of $\tau_{W}$ for each line and colour coded with $R_{2-1/1-0}$ the same colour scale as Fig. \ref{fig:quantities}} 
\label{fig:taus}
\end{figure}

To investigate the influence that line opacity has on the value of $R_{2-1/1-0}$, we have computed the optical depths of both CO lines using {\sc radmc-3d}. For each line of sight, we first compute the optical depth individually for each velocity channel. We then average these values to produce a single representative value of $\tau$. To construct this average, we weight the contribution of each velocity channel by the contribution it makes to the velocity-integrated brightness temperature of the line, i.e.

\begin{equation}
\tau_{\rm W} = \frac{\sum T_{{\rm b}, i} \times \tau_i \times {\rm d}v}{W_{\rm CO}}
\end{equation} 
where $T_{{\rm b}, i}$ is the brightness temperature in velocity channel $i$, $\tau_i$ is the corresponding optical depth, and ${\rm d}v$ is the width of the channel. It is worth noting that the way $\tau_i$ is computed is analogous to how an image is computed in {\sc radmc-3d} \citep{2012ascl.soft02015D}. Therefore $\tau_i$ is not the mean opacity seen by a single cell along the line of sight, but rather the total integrated opacity along the line of sight for each velocity channel. As such the resulting image is dependent on the 3D velocity field, in the same way our integrated intensity maps are, given that we use the LVG approximation to account for non-LTE effects.

In Fig.~\ref{fig:taus}, panels (i) and (ii) show how $\tau_{\rm W}$ varies as a function of position for the CO 1-0 and 2-1 lines, respectively. A quick comparison with Fig.~\ref{fig:MAPS} suggests a correlation between optically thick lines of sight ($\tau > 1$) and  $R_{2-1/1-0} \sim 0.7$. Additionally, along these lines of sight, $\tau_{W21}$ seems to both be larger and to increase faster than $\tau_{W10}$. 

The correlation between $\tau$ and $R_{2-1/1-0}$ becomes more evident in panels (iii) and (iv) of Fig.~\ref{fig:taus}. Values of $R_{2-1/1-0} \sim 0.7$ can be mostly attributed to emission coming from optically thick lines of sight, whereas values of $R_{2-1/1-0} \sim 0.3$ originate from optically thin lines of sight. The influence of $\tau$ on the level populations is further emphasised in panels (v) and (vi), where we plot the ratio of the CO excitation temperature $T_{\rm ex}$ and the kinetic temperature $T_{\rm kin}$ as a function of $\tau_{\rm W}$. Panel (v) shows the results for the 1-0 line and panel (vi) shows the results for the 2-1 line. We see that along most lines of sight with $\tau < 1$, both transitions are strongly sub-thermal, with excitation temperatures that are much less than the kinetic temperature. This is to be expected: we have already seen that most of this emission comes from gas with a density $n < 100 \: {\rm cm^{-3}}$, far below the critical density of even the 1-0 line, and since $\tau < 1$, even if a small amount of radiative trapping occurs, it is insufficient to change this conclusion. 

Along lines of sight with $\tau > 1$, the behaviour is more complex. On physical grounds, we expect that $T_{\rm ex} \rightarrow T$ as $\tau \rightarrow \infty$, where T is the kinetic temperature of the gas. From Fig.~\ref{fig:taus}, we see that we do indeed recover this behaviour for some of the gas along the sight-lines with high optical depth. However, we also see that there are other regions along each high $\tau$ sight-line where the emission is sub-thermal. The key to understanding this behaviour is the fact that the quantity directly responsible for influencing the level populations is not the same quantity that we are dealing with when we compute and plot $\tau_{\rm W}$. Although there is a correlation between the mean optical depth along a sight-line and the angle-averaged optical depths probed by that sight-line, the highly inhomogeneous structure of the cloud causes this correlation to be fairly weak \citep[see e.g.][]{2014MNRAS.444.2396C}. 
Therefore, the emission that we see along the $\tau > 1$ sight-lines comes from a mix of sub-thermal and thermalised gas, explaining why we do not simply recover a value of 1 for $R_{2-1/1-0}$.

\section{Discussion}\label{dissec}
Our synthetic images recover the expected peak in the CO 2-1/1-0 line ratio at $R_{2-1/1-0} \sim 0.7$, but also indicate the existence of a second peak in the line ratio distribution at $R_{2-1/1-0} \sim 0.3$. The first peak is produced by emission from CO in cold, dense regions of the cloud and our analysis in the previous section shows that much of this CO is optically thick. On the other hand, the second peak is produced by emission from optically thin, sub-thermally excited CO in warm, diffuse regions of the cloud. 

Our analysis therefore suggests that the value of $R_{2-1/1-0}$ can potentially be used as a probe of the physical conditions within a molecular cloud. By detecting both CO lines and determining whether $R_{2-1/1-0}$ is found in the high peak or the low peak, we can place constraints on the density, temperature and optical depth of the CO-emitting gas. Validating the existence of the low ratio peak with observational studies will therefore be important to establish $R_{2-1/1-0}$ as a potential observational tool for confidently distinguishing the different regions within a GMC. However, we caution that it remains to be seen whether the $R_{2-1/1-0}$ distribution that we see in this particular cloud is universal or is a result of our choice of initial conditions and/or ISRF. However, what is clear is that for this particular cloud -- i.e. the physical properties, and environmental conditions -- $R_{2-1/1-0}$ has a well-defined bimodal structure that corresponds to the physical state of the gas, such as its temperature, density and resulting level populations.  

\section{Conclusions}\label{consec}

We have used a numerical simulation of a turbulent molecular cloud to investigate the behaviour of the ratio of the velocity-integrated brightness temperatures of the first two emission lines of CO, defined as $R_{2-1/1-0}=W_{21}/W_{10}$. Our simulated cloud has properties similar to those found in nearby star-forming clouds. We have used SPH to model the chemical, thermal and dynamical evolution of the cloud, and then post-processed the simulation output using a radiative transfer code to generate synthetic CO emission maps. Our main findings can be summarised as follows:
\begin{enumerate}
\item The area-weighted PDF of $R_{2-1/1-0}$ has a bimodal distribution with two main peaks, at $\sim 0.7$ and $\sim0.3$. This clear bimodal structure correlates well with the optical depths of the CO lines. Along optically thin lines of sight, the CO excitation is strongly sub-thermal and the resulting value of  $R_{2-1/1-0}$ is small. On the other hand, along optically thick lines of sight, we probe a mix of sub-thermal and thermal emission, resulting in a much higher value of the line ratio.
\item The high ratio peak primarily traces the cold ($T \leq 40$~K) and dense ($n \geq 10^3 \: {\rm cm^{-3}}$) molecular gas within the molecular cloud. This value is similar to the ``canonical'' value of $R_{2-1/1-0}$ often quoted in the literature, and also to the values measured in local molecular clouds. \citep{1994ApJ...425..641S,2015ApJS..216...18N} 
\item The low ratio peak traces more diffuse ($n \leq 10^3 \: {\rm cm^{-3}}$) and warmer ($T \geq 40$~K) molecular gas within the cloud. This gas contains much less CO and so the
emission from these regions is much fainter, requiring high sensitivity to detect. We note that  \citet{2015ApJS..216...18N} reported values of $R_{2-1/1-0} \sim 0.4-0.5$ towards the outskirts of Orion, consistent with the range of values we find for their limiting sensitivity.

\end{enumerate}

As such the value of $R_{2-1/1-0}$ can be indicative of the physical conditions in a particular region of a cloud. Further study, exploring a wide range of environmental conditions, is required to see whether the result we present here is universal. We will follow up on this in a future paper.

\section*{Acknowledgements}
We thank the anonymous referee for a constructive report that improved the manuscript. We would also like to thank Cornelis Dullemond for his help with the RADMC-3D refinement.  PCC acknowledges support from the Science and Technology Facilities Council (under grant ST/N00706/1) and the European Community's Horizon 2020 Programme H2020-COMPET-2015, through the StarFormMapper project (number 687528).  SCOG and RSK acknowledge financial support from the Deutsche Forschungsgemeinschaft  via SFB 881, ``The Milky Way System'' (sub-projects B1, B2 and B8) and SPP 1573, ``Physics of the Interstellar Medium''. They also acknowledge support from the European Research Council under the European Community's Seventh Framework Programme (FP7/2007-2013) via the ERC Advanced Grant STARLIGHT (project number 339177).

\bibliography{Bibliography}

\begin{thebibliography}{52}
\expandafter\ifx\csname natexlab\endcsname\relax\def\natexlab#1{#1}\fi

\bibitem[{{Barriault}, {Joncas} \& {Plume}(2011){Barriault}, {Joncas}, \&
  {Plume}}]{2011MNRAS.416.1250B}
{Barriault} L., {Joncas} G., {Plume} R., 2011, \mnras, 416, 1250

\bibitem[{{Bigiel} {et~al}\mbox{.}(2008){Bigiel}, {Leroy}, {Walter}, {Brinks},
  {de Blok}, {Madore}, \& {Thornley}}]{2008AJ....136.2846B}
{Bigiel} F., {Leroy} A., {Walter} F., {Brinks} E., {de Blok} W.~J.~G., {Madore}
  B., {Thornley} M.~D., 2008, \aj, 136, 2846

\bibitem[{Bisbas {et~al}\mbox{.}(2012)Bisbas, Bell, Viti, Yates, \&
  Barlow}]{2012MNRAS.427.2100B}
Bisbas T.~G., Bell T.~A., Viti S., Yates J., Barlow M.~J., 2012, MNRAS, 427,
  2100

\bibitem[{{Black}(1994)}]{Black1994}
{Black} J.~H., 1994, in Astronomical Society of the Pacific Conference Series,
  Vol.~58, The First Symposium on the Infrared Cirrus and Diffuse Interstellar
  Clouds, {Cutri} R.~M., {Latter} W.~B., eds., p. 355

\bibitem[{Bolatto, Wolfire \& Leroy(2013)Bolatto, Wolfire, \&
  Leroy}]{Bolatto:2013hl}
Bolatto A.~D., Wolfire M., Leroy A.~K., 2013, \araa, 51, 207

\bibitem[{{Brand} \& {Wouterloot}(1995)}]{1995A&A...303..851B}
{Brand} J., {Wouterloot} J.~G.~A., 1995, \aap, 303, 851

\bibitem[{{Casoli} {et~al}\mbox{.}(1991){Casoli}, {Dupraz}, {Combes}, \&
  {Kazes}}]{1991A&A...251....1C}
{Casoli} F., {Dupraz} C., {Combes} F., {Kazes} I., 1991, \aap, 251, 1

\bibitem[{Castets {et~al}\mbox{.}(1990)Castets, Duvert, Dutrey, Bally, Langer,
  \& Wilson}]{1990A&A...234..469C}
Castets A., Duvert G., Dutrey A., Bally J., Langer W.~D., Wilson R.~W., 1990,
  A{\&}A, 234, 469

\bibitem[{{Clark} \& {Glover}(2014)}]{2014MNRAS.444.2396C}
{Clark} P.~C., {Glover} S.~C.~O., 2014, \mnras, 444, 2396

\bibitem[{Clark \& Glover(2015)}]{2015MNRAS.452.2057C}
Clark P.~C., Glover S. C.~O., 2015, \mnras, 452, 2057

\bibitem[{{Clark}, {Glover} \& {Klessen}(2012a){Clark}, {Glover}, \&
  {Klessen}}]{Clark_etal_2012}
{Clark} P.~C., {Glover} S.~C.~O., {Klessen} R.~S., 2012a, \mnras, 420, 745

\bibitem[{{Clark} {et~al}\mbox{.}(2012b){Clark}, {Glover}, {Klessen}, \&
  {Bonnell}}]{Clark2012b}
{Clark} P.~C., {Glover} S.~C.~O., {Klessen} R.~S., {Bonnell} I.~A., 2012b,
  \mnras, 424, 2599

\bibitem[{Dame, Hartmann \& Thaddeus(2001)Dame, Hartmann, \&
  Thaddeus}]{Dame:2001bg}
Dame T.~M., Hartmann D., Thaddeus P., 2001, ApJ, 547, 792

\bibitem[{{Draine}(1978)}]{Draine1978}
{Draine} B.~T., 1978, \apjs, 36, 595

\bibitem[{Dullemond(2012)}]{2012ascl.soft02015D}
Dullemond C.~P., 2012, ASCL, 1202.015

\bibitem[{{Eckart} {et~al}\mbox{.}(1990){Eckart}, {Downes}, {Genzel}, {Harris},
  {Jaffe}, \& {Wild}}]{1990ApJ...348..434E}
{Eckart} A., {Downes} D., {Genzel} R., {Harris} A.~I., {Jaffe} D.~T., {Wild}
  W., 1990, \apj, 348, 434

\bibitem[{{Field}(1965)}]{1965ApJ...142..531F}
{Field} G.~B., 1965, \apj, 142, 531

\bibitem[{{Field}, {Goldsmith} \& {Habing}(1969){Field}, {Goldsmith}, \&
  {Habing}}]{1969ApJ...155L.149F}
{Field} G.~B., {Goldsmith} D.~W., {Habing} H.~J., 1969, \apjl, 155, L149

\bibitem[{{Fryxell} {et~al}\mbox{.}(2000){Fryxell}, {Olson}, {Ricker},
  {Timmes}, {Zingale}, {Lamb}, {MacNeice}, {Rosner}, {Truran}, \&
  {Tufo}}]{Fryxell00}
{Fryxell} B. {et~al.}, 2000, \apjs, 131, 273

\bibitem[{Glover \& Clark(2012{\natexlab{a}})}]{Glover:2012et}
Glover S. C.~O., Clark P.~C., 2012{\natexlab{a}}, MNRAS, 421, 116

\bibitem[{Glover \& Clark(2012{\natexlab{b}})}]{Glover:2012dd}
Glover S. C.~O., Clark P.~C., 2012{\natexlab{b}}, MNRAS, 421, 9

\bibitem[{Glover \& Clark(2012{\natexlab{c}})}]{Glover:2012jo}
Glover S. C.~O., Clark P.~C., 2012{\natexlab{c}}, MNRAS, 426, 377

\bibitem[{{Glover} {et~al}\mbox{.}(2015){Glover}, {Clark}, {Micic}, \&
  {Molina}}]{Glover15}
{Glover} S.~C.~O., {Clark} P.~C., {Micic} M., {Molina} F., 2015, \mnras, 448,
  1607

\bibitem[{Glover {et~al}\mbox{.}(2010)Glover, Federrath, Mac~Low, \&
  Klessen}]{Glover:2010bu}
Glover S. C.~O., Federrath C., Mac~Low M.-M., Klessen R.~S., 2010, MNRAS, 404,
  2

\bibitem[{Glover \& Mac~Low(2007{\natexlab{a}})}]{Glover:2007gr}
Glover S. C.~O., Mac~Low M.-M., 2007{\natexlab{a}}, ApJS, 169, 239

\bibitem[{Glover \& Mac~Low(2007{\natexlab{b}})}]{Glover:2007wq}
Glover S. C.~O., Mac~Low M.-M., 2007{\natexlab{b}}, ApJ, 659, 1317

\bibitem[{Goldsmith {et~al}\mbox{.}(2008)Goldsmith, Heyer, Narayanan, Snell,
  Li, \& Brunt}]{2008ApJ...680..428G}
Goldsmith P.~F., Heyer M., Narayanan G., Snell R., Li D., Brunt C., 2008, ApJ,
  680, 428

\bibitem[{{Habing}(1968)}]{Habing1968}
{Habing} H.~J., 1968, \bain, 19, 421

\bibitem[{{Hasegawa}(1997)}]{1997IAUS..170...39H}
{Hasegawa} T., 1997, in IAU Symposium, Vol. 170, IAU Symposium, {Latter} W.~B.,
  {Radford} S.~J.~E., {Jewell} P.~R., {Mangum} J.~G., {Bally} J., eds., pp.
  39--46

\bibitem[{{Klessen} \& {Glover}(2016)}]{2016SAAS...43...85K}
{Klessen} R.~S., {Glover} S.~C.~O., 2016, Star Formation in Galaxy Evolution:
  Connecting Numerical Models to Reality, Saas-Fee Advanced Course, Volume
  43.~ISBN 978-3-662-47889-9.~Springer-Verlag Berlin Heidelberg, 2016, p.~85,
  43, 85

\bibitem[{Koda {et~al}\mbox{.}(2012)Koda, Scoville, Hasegawa, Calzetti,
  Donovan~Meyer, Egusa, Kennicutt, Kuno, Louie, Momose, Sawada, Sorai, \&
  Umei}]{Koda:2012ht}
Koda J. {et~al.}, 2012, ApJ, 761, 41

\bibitem[{Larson(1981)}]{1981MNRAS.194..809L}
Larson R.~B., 1981, MNRAS, 194, 809

\bibitem[{{Leroy} {et~al}\mbox{.}(2009){Leroy}, {Walter}, {Bigiel}, {Usero},
  {Weiss}, {Brinks}, {de Blok}, {Kennicutt}, {Schuster}, {Kramer},
  {Wiesemeyer}, \& {Roussel}}]{2009AJ....137.4670L}
{Leroy} A.~K. {et~al.}, 2009, \aj, 137, 4670

\bibitem[{Liszt \& Lucas(1998)}]{Liszt:1998tx}
Liszt H.~S., Lucas R., 1998, A{\&}A, 339, 561

\bibitem[{Liszt \& Pety(2012)}]{2012A&A...541A..58L}
Liszt H.~S., Pety J., 2012, A{\&}A, 541, A58

\bibitem[{{Maddalena} {et~al}\mbox{.}(1986){Maddalena}, {Morris}, {Moscowitz},
  \& {Thaddeus}}]{Maddalena:1986}
{Maddalena} R.~J., {Morris} M., {Moscowitz} J., {Thaddeus} P., 1986, \apj, 303,
  375

\bibitem[{McKee \& Ostriker(2007)}]{McKee:2007bd}
McKee C.~F., Ostriker E.~C., 2007, \araa, 45, 565

\bibitem[{Nelson \& Langer(1999)}]{NelsonLanger1999}
Nelson R.~P., Langer W.~D., 1999, ApJ, 524, 923

\bibitem[{Nishimura {et~al}\mbox{.}(2015)Nishimura, Tokuda, Kimura, Muraoka,
  Maezawa, Ogawa, Dobashi, Shimoikura, Mizuno, Fukui, \&
  Onishi}]{2015ApJS..216...18N}
Nishimura A. {et~al.}, 2015, ApJS, 216, 18

\bibitem[{Offner {et~al}\mbox{.}(2013)Offner, Bisbas, Viti, \&
  Bell}]{Offner:2013du}
Offner S. S.~R., Bisbas T.~G., Viti S., Bell T.~A., 2013, ApJ, 770, 49

\bibitem[{{Price}(2012)}]{2012JCoPh.231..759P}
{Price} D.~J., 2012, Journal of Computational Physics, 231, 759

\bibitem[{{Sakamoto} {et~al}\mbox{.}(1997){Sakamoto}, {Hasegawa}, {Handa},
  {Hayashi}, \& {Oka}}]{1997ApJ...486..276S}
{Sakamoto} S., {Hasegawa} T., {Handa} T., {Hayashi} M., {Oka} T., 1997, \apj,
  486, 276

\bibitem[{Sakamoto {et~al}\mbox{.}(1994)Sakamoto, Hayashi, Hasegawa, Handa, \&
  Oka}]{1994ApJ...425..641S}
Sakamoto S., Hayashi M., Hasegawa T., Handa T., Oka T., 1994, ApJ, 425, 641

\bibitem[{{Sawada} {et~al}\mbox{.}(2001){Sawada}, {Hasegawa}, {Handa},
  {Morino}, {Oka}, {Booth}, {Bronfman}, {Hayashi}, {Luna Castellanos}, {Nyman},
  {Sakamoto}, {Seta}, {Shaver}, {Sorai}, \& {Usuda}}]{2001ApJS..136..189S}
{Sawada} T. {et~al.}, 2001, \apjs, 136, 189

\bibitem[{Sch{\"o}ier {et~al}\mbox{.}(2005)Sch{\"o}ier, van~der Tak, van
  Dishoeck, \& Black}]{2005A&A...432..369S}
Sch{\"o}ier F.~L., van~der Tak F. F.~S., van Dishoeck E.~F., Black J.~H., 2005,
  A{\&}A, 432, 369

\bibitem[{Shetty {et~al}\mbox{.}(2011)Shetty, Glover, Dullemond, \&
  Klessen}]{Shetty:2011eh}
Shetty R., Glover S.~C., Dullemond C.~P., Klessen R.~S., 2011, MNRAS, 412, 1686

\bibitem[{Sobolev(1957)}]{1957SvA.....1..678S}
Sobolev V.~V., 1957, SOVAST, 1, 678

\bibitem[{Springel(2005)}]{Springel:2005cz}
Springel V., 2005, MNRAS, 364, 1105

\bibitem[{{Sz{\H u}cs}, {Glover} \& {Klessen}(2016){Sz{\H u}cs}, {Glover}, \&
  {Klessen}}]{2016MNRAS.460...82S}
{Sz{\H u}cs} L., {Glover} S.~C.~O., {Klessen} R.~S., 2016, \mnras, 460, 82

\bibitem[{Sz{\H{u}}cs, Glover \& Klessen(2014)Sz{\H{u}}cs, Glover, \&
  Klessen}]{2014MNRAS.445.4055S}
Sz{\H{u}}cs L., Glover S. C.~O., Klessen R.~S., 2014, MNRAS, 445, 4055

\bibitem[{van Dishoeck \& Black(1988)}]{1988ApJ...334..771V}
van Dishoeck E.~F., Black J.~H., 1988, ApJ, 334, 771

\bibitem[{Yang {et~al}\mbox{.}(2010)Yang, {Stancil, P. C.}, Balakrishnan, \&
  Forrey}]{2010ApJ...718.1062Y}
Yang B., {Stancil, P. C.}, Balakrishnan N., Forrey R.~C., 2010, ApJ, 718, 1062

\end{thebibliography}
\bibliographystyle{mn2e}

\appendix
\section{Achieving convergence} \label{appsec}
As mentioned in Section \ref{gridsubsec}, to carry out radiative transfer simulations with RADMC-3D requires us to map the unstructured SPH particle distribution to a Cartesian grid. Interpolation onto a uniform grid is relatively straightforward, but such grids are not a good match for the highly non-uniform distribution of the SPH particles, with the result that high resolution is required to achieve convergence even for the CO 1-0 line \citep{Glover15,2016MNRAS.460...82S}. Achieving convergence for the higher $J$ transitions is even more difficult, owing to their higher critical densities, and requires an unfeasibly large uniform grid. To avoid this problem, we have implemented a module within RADMC-3D that allows snapshots from GADGET-2 to be readily interpolated onto a hierarchically-structured oct-tree grid. In the remainder of this section, we discuss how we construct this grid and carry out the interpolation onto it. 

\subsection*{Grid construction}
We construct our grid iteratively, using a method similar to that used in adaptive mesh refinement (AMR) simulations of hydrodynamical flows. We begin with a uniform Cartesian base grid with a specified resolution. We then loop over the full set of SPH particles. For each SPH particle, we first identify the cell in our grid that contains the particle, and next apply a refinement criterion to determine whether or not the grid cell needs to be refined in order to properly represent the particle. If the cell needs to be refined, we split it into 8 sub-cells, each occupying $1/8^{\rm th}$ of the volume of the parent cell. This process is then repeated for the same particle, using the refined grid, until the cell in which the particle is located satisfies the refinement criterion. 

To determine whether or not a given cell needs to be refined, we compare a measure of its size, $d$, with the smoothing length of the particle, $h$. We define $d$ using the equation
\begin{equation}
d = \sqrt{(\Delta x)^{2} + (\Delta y)^{2} + (\Delta z)^{2}},
\end{equation}
where $\Delta x$, $\Delta y$, $\Delta z$ are the lengths of the side of the cell in the $x, y$, and $z$ directions, respectively. Note that for a cubical base grid with the same initial resolution in each dimension, $\Delta x = \Delta y = \Delta z$ and this simplifies to $d = \sqrt{3} \Delta x$. We have examined a number of different refinement criteria and we find that the one which gives us the required resolution with the least computational overhead is the requirement that $d \leq h/2$.

\begin{figure}
\includegraphics[width=\linewidth]{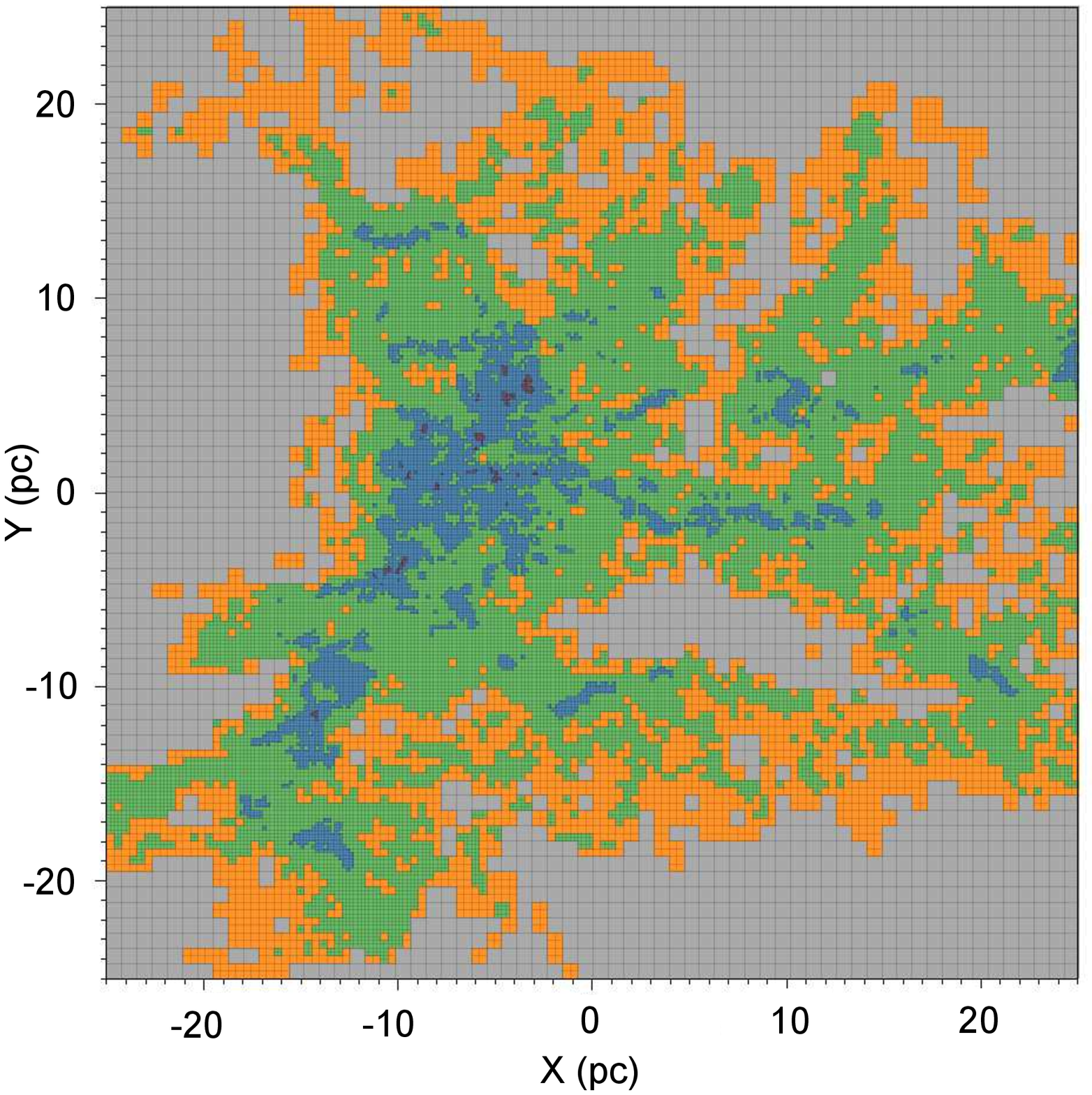}
\caption {A slice through the unstructured AMR grid created with our refinement algorithm. Each colour corresponds to a different, more refined layer of the grid}
\label{fig:refinement}
\end{figure}

Once the grid satisfies the refinement criterion for the considered SPH particle, we then proceed to the next particle in the list and apply the same procedure. We continue in this way until we have looped over the full set of SPH particles. The structure of the resulting grid closely resembles the structure of the SPH particle distribution, as illustrated in Fig.~\ref{fig:refinement}. Finally, at the end of the refinement procedure, we loop one further time over all particles, verifying that each particle is associated with a grid cell, and carrying out the interpolation from the particles to the grid, as described below. 

When constructing the emission maps analysed in this paper, we have used a base grid of $64^{3}$ cells. After refinement, using the refinement criterion $d \leq h/2$, we find that the mostly highly-refined cells have been refined a total of six times, resulting in a local resolution equivalent to what one would achieve using a $4096^3$ uniform grid. However, the total number of grid cells in the refined grid is only $\sim 145^{3}$. Our refinement procedure therefore allows us to reduce the memory requirements and computational cost by a factor of more than $2 \times 10^{4}$.

\subsection*{Interpolation}

We use standard SPH kernel interpolation to interpolate the CO and H$_{2}$ number densities, the gas temperature etc.\ from our particle distribution onto the oct-tree grid. Briefly, if we have some scalar quantity $A$ that is a function of position, then its value at position ${\mathbf r}$ is given by the sum \citep{2012JCoPh.231..759P}:
\begin{equation}
A({\mathbf r}) \simeq \sum_{b=1}^{N_{n}} m_{b} \frac{A_{b}}{\rho_{b}} W({\mathbf r} - {\mathbf r_{b}}, h_{b}), 
\end{equation}
where $m_{b}$ is the mass of SPH particle $b$, $\rho_{b}$ and $A_{b}$ are the values of the density and the scalar $A$ carried by particle $b$, ${\mathbf r}_{b}$ is that particle's position, $W$ is the SPH smoothing kernel, $h_{b}$ is the smoothing length of particle $b$, and we sum over all $N_{n}$ particles for which $W > 0$. GADGET-2 uses a cubic spline kernel with 
\begin{equation}
W(r,h) = \frac{8}{\pi h^3} \times \left\{ 
\begin{array}{l l}
1 - 6\big(\frac{r}{h}\big)^2 + 6\big(\frac{r}{h}\big)^3 & \textrm{if $0 \leq \frac{r}{h} \leq 0.5$}\\
2\big(1- \frac{r}{h}\big)^3 & \textrm{if $0.5 \leq \frac{r}{h} \leq 1$}\\
0 & \textrm{if $\frac{r}{h} > 1$},		
\end{array}
\right.
\end{equation}
and so for any given cell within the oct-tree grid, the only particles which contribute are those for which the smoothing length of the particle is larger than the distance from the particle to the centre of the cell. The refinement procedure described above ensures that for each SPH particle, there is at least one cell for which $r < h$, and so none of the information from the particles is lost. 

\subsection*{Reaching Convergence}

\begin{figure}
\includegraphics[width=\linewidth]{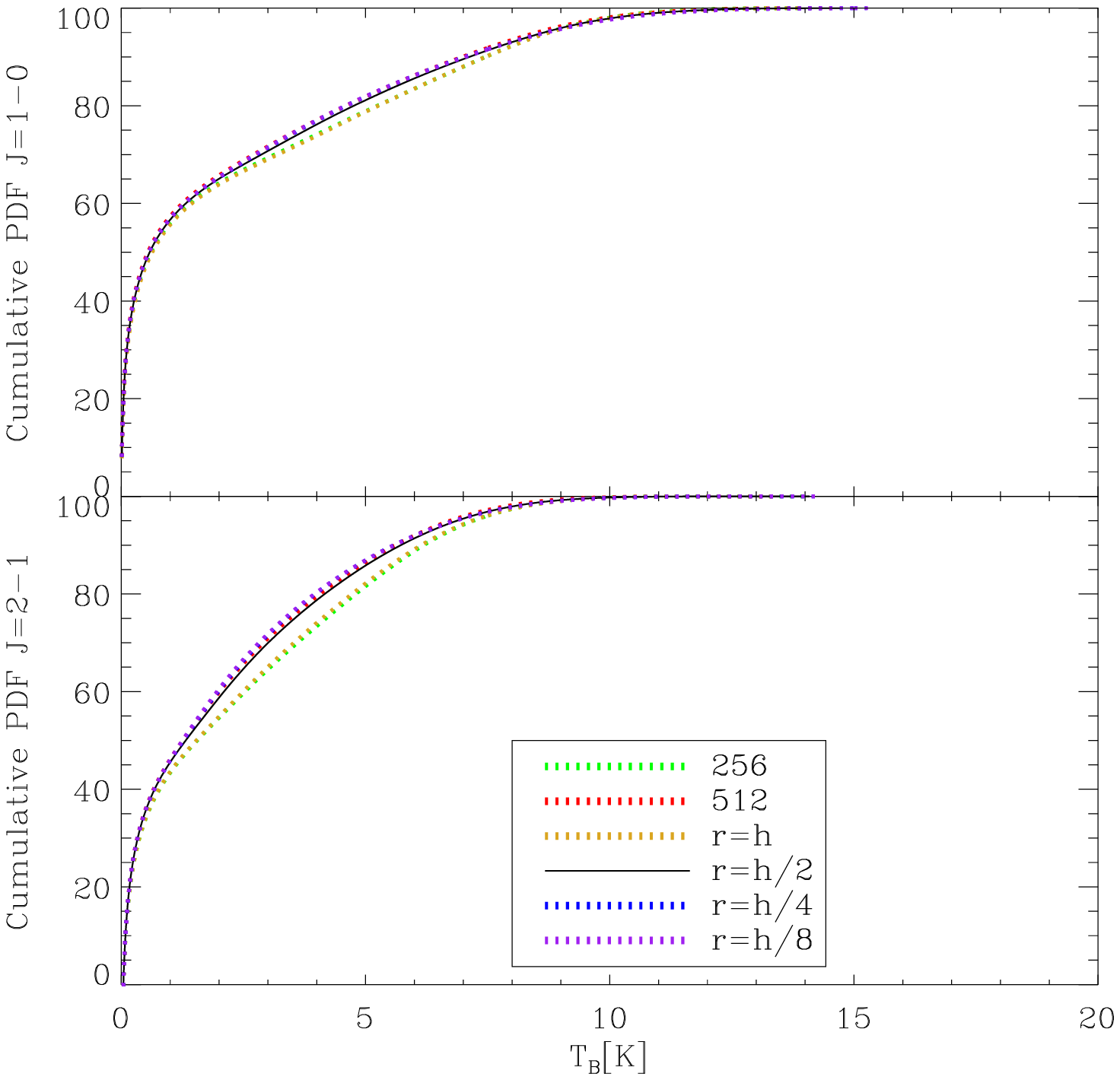}
\caption {The cumulative luminosity from the PPV cube plotted against the brightness temperature of CO 1-0 (top panel) and CO 2-1 (bottom panel). The black solid line represents the resolution used for this study. The other lines correspond to different grid resolutions: the green dotted and red dotted lines correspond to fixed grids with resolutions of $256^{3}$ and $512^{3}$ respectively, while the remaining three lines correspond to different choices for the refinement criterion in the adaptive approach.}
\label{fig:Res_Tb}
\end{figure}

\begin{figure}
\includegraphics[width=\linewidth]{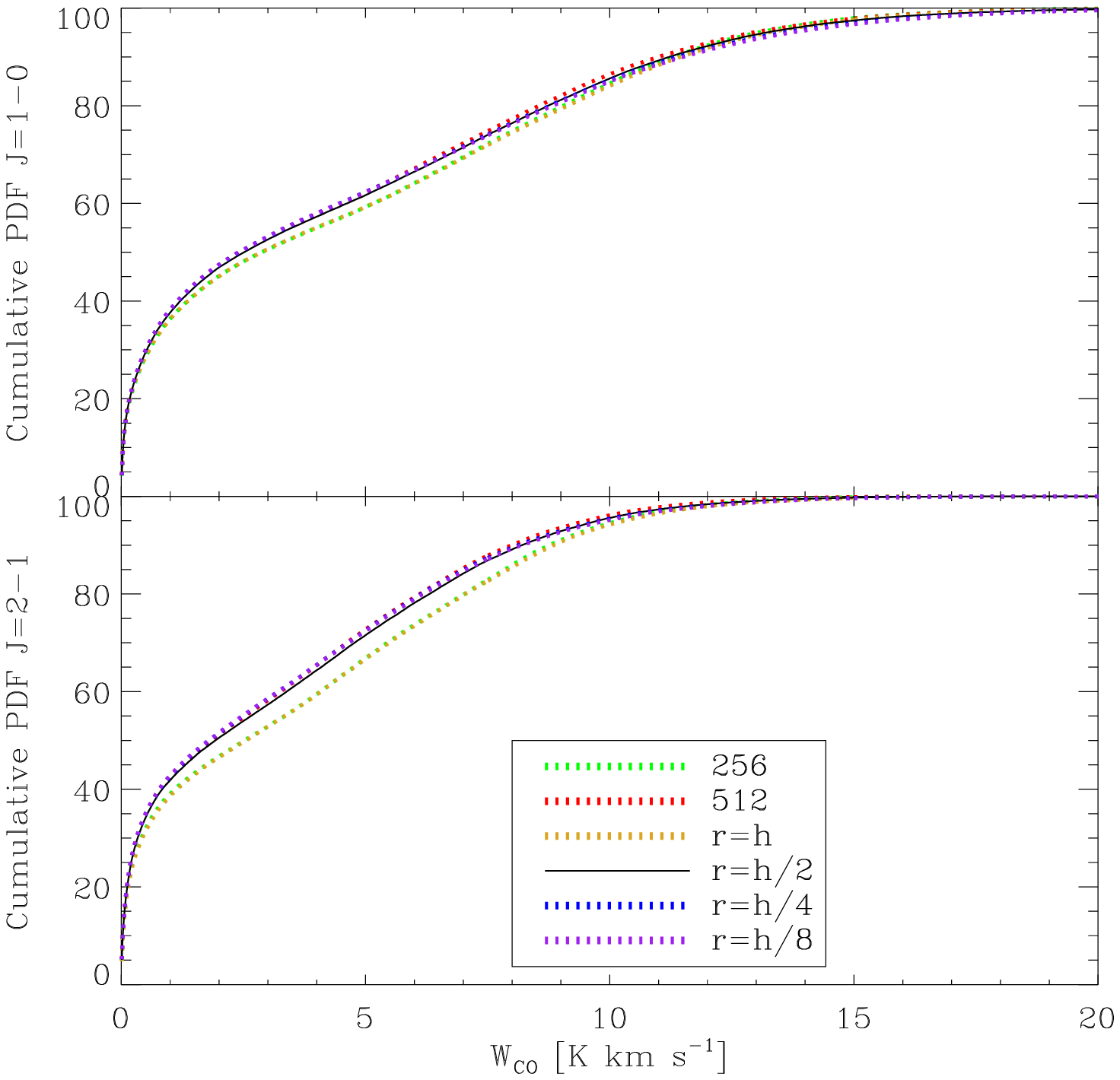}
\caption {Similar plot to Fig.~\ref{fig:Res_Tb}, however in this case the cumulative luminosity of the integrated intensity map is plotted against integrated intensity. }
\label{fig:Res_WCO}
\end{figure}

To test that our simulations are converged we ran a new set of simulations in which the only parameter varied was resolution. These results are plotted in Figs.~\ref{fig:Res_Tb} and~\ref{fig:Res_WCO}.

Fig.~\ref{fig:Res_Tb} shows the cumulative CO emission in PPV space plotted against brightness temperature $T_B$. This is done for CO 1-0 and CO 2-1 for every resolution tested. Fig.~\ref{fig:Res_WCO} shows a similar plot for the cumulative CO emission in the integrated intensity map. A fixed resolution of $256^3$ and an AMR grid with refinement criterion of $r=h$ do not reach convergence since the emission from low intensity regions is underestimated. On the other hand the cumulative emission curves for a fixed resolution of $512^3$ and refinement criterions above $r=h/2$ are quite similar, which suggests that these are converged. 

In this case a $512^3$ resolution seems to be enough to resolve most of the particles in from the SPH snapshot and therefore provide convergence. However it is worth pointing out that this is only the case due to the nature of the studied cloud, i.e.\ a small, low density cloud. Nonetheless at a fourth of the computational cost, our AMR refinement method with a refinement criterion of $r=h/2$ proves to be better.

\end{document}